\documentclass[a4paper,11pt]{article}
\usepackage[top=1in, bottom=1.25in, left=1.25in, right=1.25in]{geometry}

\usepackage{amsmath}
\usepackage{amssymb}

\usepackage{graphicx}
\usepackage{subfigure}
\usepackage{setspace}

\usepackage{hyperref}

\usepackage{natbib}

\title{Deep learning electromagnetic inversion with convolutional neural networks}

\author{Vladimir Puzyrev\thanks{School of Earth and Planetary Sciences and Curtin Oil and Gas Innovation Centre, Curtin University, Perth, Australia \tt \href{mailto:vladimir.puzyrev@gmail.com}{vladimir.puzyrev@gmail.com}} }

\date{December 26, 2018}

\begin{document}

\label{firstpage}

\maketitle

\begin{abstract}
Geophysical inversion attempts to estimate the distribution of physical properties in the Earth's interior from observations collected at or above the surface. Inverse problems are commonly posed as least-squares optimization problems in high-dimensional parameter spaces. Existing approaches are largely based on deterministic gradient-based methods, which are limited by nonlinearity and nonuniqueness of the inverse problem. Probabilistic inversion methods, despite their great potential in uncertainty quantification, still remain a formidable computational task. In this paper, I explore the potential of deep learning methods for electromagnetic inversion. This approach does not require calculation of the gradient and provides results instantaneously. Deep neural networks based on fully convolutional architecture are trained on large synthetic datasets obtained by full 3-D simulations. The performance of the method is demonstrated on models of strong practical relevance representing an onshore controlled source electromagnetic $\text{CO}_\text{2}$ monitoring scenario. The pre-trained networks can reliably estimate the position and lateral dimensions of the anomalies, as well as their resistivity properties. Several fully convolutional network architectures are compared in terms of their accuracy, generalization, and cost of training. Examples with different survey geometry and noise levels confirm the feasibility of the deep learning inversion, opening the possibility to estimate the subsurface resistivity distribution in real time.
\end{abstract}

\section{Introduction}

Electrical resistivity is an important property of geological formations with high sensitivity to porosity and fluid saturation. Electromagnetic (EM) methods are widely applied in various areas of geophysics including hydrocarbon and mineral exploration, $\text{CO}_\text{2}$ storage monitoring, geothermal reservoir characterization, and hydrogeological studies. These methods utilize low- and medium-frequency EM energy to map variations in the subsurface resistivity and characterize the structure of the geological formations at depths ranging from a few meters to tens of kilometers. The interpretation of EM measurements collected in complex geological settings typically relies on inversion when a subsurface model described by a set of parameters (e.g., isotropic or anisotropic resistivity) is progressively updated by minimizing the discrepancies between the observations and the model predictions until they fall below a chosen tolerance.

Inverse problems are not restricted to geophysics and arise in many fields including medical imaging and nondestructive testing. The difficulties in solving them come from several constraints imposed by real-world scenarios. Nearly all geophysical inverse problems are notoriously nonunique (many models fit the data equally well) and nonlinear (small variations in the measured data may induce large variations in the model). The methods for solving inverse problems can be classified into two main categories: local and global optimization algorithms \citep{nocedal2006nonlinear}. Local methods seek only a local solution, a model for which the misfit is smaller than for all other ``nearby'' models. This search is typically performed in an iterative way and involves lengthy calculations. For convex problems, local solutions are also global solutions; however, geophysical inverse problems are highly nonlinear and in practice, local optimization methods are not able to find the global solution. Global methods, in turn, are very expensive in terms of computation power and time, and this grows exponentially as the number of model parameters increases. Another major obstacle is that the acquired data typically contains a high level of background geological noise \citep{everett2002geological} and often lacks full coverage as a result of equipment and data collection constraints, which also increases the nonuniqueness of the solution. Combining several different EM techniques or using them together with seismic methods that resolve geological structures with much higher resolution brings additional challenges.

Inversion of three-dimensional (3-D) data is a computationally challenging task, whose cost is mainly due to the solution of multiple forward modelling problems. Modern seismic and EM surveys include hundreds and thousands of sources and receivers, thus significantly increasing the computational cost of forward modelling. Accurate simulations of EM phenomena require fine-scale models and lead to forward problems with tens of millions of unknowns \citep{shantsev2017large}. Consequently, inversion of a large geophysical dataset using traditional deterministic approaches requires the solution of several millions of forward problems each having up to tens of millions of unknowns. Two local optimization algorithms commonly used in large-scale inversion of EM data are the nonlinear conjugate gradient (NLCG) method \citep{rodi2001nonlinear, commer2008new, moorkamp2011framework} and the limited memory Broyden-Fletcher-Goldfarb-Shanno ($l$-BFGS) algorithm \citep{haber2004quasi, plessix2008resistivity, avdeev20093d}. The latter belongs to a family of quasi-Newton methods and approximates the Hessian with past gradients. Second-order optimization methods were applied to 2-D problems in the past \citep{abubakar2006two} and have become common in 3-D in recent years \citep{grayver2013three, amaya2016low, nguyen2016comparing}. One of the main practical limitations of these algorithms is their high memory cost. The size of realistic 3-D datasets has often been considered excessively large for Gauss-Newton inversion, thereby preventing the widespread use of second-order algorithms in industry-scale controlled source electromagnetic (CSEM) surveys. Probabilistic (Bayesian) inversion methods are well suited to quantify large datasets and estimate uncertainties \citep{sambridge2002monte, sen2013global}. However, the computational cost of probabilistic analysis is notoriously high, particularly in high-dimensional parameter spaces. The real-time processing of acquired measurements requires some simplifications to the model or massive computational resources, often not available in field settings. All this motivates the development and practical use of new inversion methods that have modest computational demands, while being robust and efficient.

Machine learning (ML) methods have been the focus of increasing attention in the geoscience community in recent years. The principal reason for this is the recent rise of deep learning (DL) in almost every field of science and engineering following the great success in computer vision tasks in the early 2010s \citep{krizhevsky2012imagenet}. In the following, I use the term ``deep learning'' to distinguish these novel methods, mainly based on deep networks, from classical ML algorithms. The idea of using a neural network (NN) to estimate the properties or spatial location of a simple target from EM data is not new \citep{poulton1992location}. However, the capabilities of the first networks were very limited and the method was only marginally applied in geophysics in the 90s and 00s. 

Extensive research in artificial intelligence over the last decades and the latest advancements in GPU computing have led to a tectonic shift in NN performance. One of the most powerful recent developments in DL are deep convolutional networks, which showed tremendous success in image classification problems and are actively extended to other fields nowadays. Deep networks having from hundreds of millions to billions of parameters trained on huge datasets have become common in many applications \citep[e.g.,][]{simonyan2014very, he2016deep, badrinarayanan2017segnet}. DL is now a crucial component of modern computer vision, speech recognition, and machine translation systems. The field has also shown great promise in approximating complex physics simulations with high degrees of precision and in orders of magnitude less time. For example, \citet{carleo2017solving} applied NNs to solve many-body problems in quantum physics. \citet{peurifoy2018nanophotonic} used DL to approximate light scattering by multilayer nanoparticles and obtained results similar to the numerical solution of Maxwell's equations orders of magnitude faster. Recent applications of DL methods in inverse problems include seismic \citep{araya2018deep} and X-ray computed tomography \citep{ye2018deep}.

The modern use of DL tools in geophysics is usually centered around the identification and classification of geological features \citep{lin2018efficient}. First studies in seismics include lithology prediction \citep{zhang2018deep}, automatic event picking \citep{chen2017automatic}, and noise removal \citep{paitz2017neural}. The first promising applications of DL in seismic modelling \citep{moseley2018fast} and tomography \citep{araya2018deep} may indicate the beginning of a new era in geophysical modelling and inversion when deep NNs will supplement or replace traditional methods in approximating physical simulations and inverting for model parameters from given observations. Inversion based on DL methods may not only yield accurate results but also produce them in real time by predicting the distribution of subsurface properties from data in a single step.

In this paper, I explore the applicability of DL methods for EM inversion and propose a scheme based on fully convolutional neural networks for 2-D inversion. The training dataset is generated by fully 3-D simulations of models with different 2-D resistivity anomalies. The remainder of the paper is organized as follows. First, I briefly describe the forward and inverse EM problems in Section 2. In Section 3, I outline the most important concepts of DL including convolutional neural networks and regularization, and discuss the loss functions employed and data augmentation procedure. The performance of the proposed method is studied in Section 4 on several examples of onshore CSEM $\text{CO}_\text{2}$ monitoring. In Section 5, I discuss the findings, implications, and limitations of DL inversion based on convolutional networks. Finally, the last section summarizes the outcomes of this study and points out future research directions.

\section{Problem formulation}

\subsection{Forward modelling}

Electromagnetic phenomena is governed by Maxwell's equations. Assuming a time harmonic dependence of the form $e^{i \omega t}$, Maxwell's equations in the frequency domain can be written as
\begin{align} \label{eq:Maxwell}
\begin{split}
& \nabla \times \mathbf{E} = - i \omega \mu \mathbf{H}, \\
& \nabla \times \mathbf{H} = \mathbf{J} + \mathbf{J}_s + i \omega \epsilon \mathbf{E} \ \ \textrm{in} \ \Omega.
\end{split}
\end{align}
Here $\mathbf{E}$ and $\mathbf{H}$ are the electric and magnetic fields, respectively, $\mathbf{J} = \sigma \textbf{E}$ is the electrical current density vector, $\mathbf{J}_s$ is the source current distribution, and $\Omega \subset \mathbb{R}^3$ is the 3-D problem domain. In the most general case, there are three medium parameters that are symmetric and positive definite 3x3 tensors: the electrical conductivity $\sigma$, the dielectric permittivity $\epsilon$, and the magnetic permeability $\mu$. Geophysical modelling codes typically deal with a magnetoquasistatic simplification of Maxwell's equations where the displacement current term $i \omega \epsilon \mathbf{E}$ is neglected due to low frequencies used and the magnetic permeability is assumed to be equal to the free space value $\mu = \mu_0$ since variations in it are rare for common geological formations. Thus, the electrical conductivity $\sigma$ (or its reciprocal, resistivity ${\rho}$) becomes the main EM characteristic of subsurface formations.

Equations \eqref{eq:Maxwell} are usually transformed to a second-order curl-curl equation for either electric or magnetic field. The most common approach is to eliminate $\mathbf{H}$ from \eqref{eq:Maxwell} and obtain the $\mathbf{E}$-field equation
\begin{equation} \label{eq:CurlCurl}
\nabla \times \frac{1}{\mu_0} \nabla \times \mathbf{E} + i \omega \sigma \mathbf{E} = - i \omega \mathbf{J_s},
\end{equation} 
or, in the scattered field formulation
\begin{equation} \label{eq:CurlCurl2}
\nabla \times \nabla \times \mathbf{E}_s + i \omega \mu_0 \sigma \mathbf{E}_s = - i \omega \mu_0 (\sigma - \sigma_p) \mathbf{E}_p,
\end{equation}
which is used to avoid excessive meshing near the sources. Here the subscript $p$ represents the background (primary) fields and medium properties, $s$ stands for the scattered (secondary) fields and $\mathbf{E} = \mathbf{E_p} + \mathbf{E_s}$. In the case of Maxwell's equations in the magnetoquasistatic formulation, the computational domain can be truncated by the perfect electrically conducting boundary conditions $\mathbf{E}_s = 0$, provided that the boundaries are located sufficiently far from the regions of interest.

The mapping from the resistivity model to the measured EM data is called the forward problem. The inverse problem discussed in the next section constitutes a nonlinear mapping from the measured field to the model. In the presented scheme, forward modelling is employed only for the generation of synthetic data used in NN training. For the spatial discretization of Equation \eqref{eq:CurlCurl2}, I use the 3-D finite-difference method based on staggered grids that is still commonly applied in geophysical problems \citep[e.g.,][]{kelbert2014modem, jaysaval2014fast, commer2015transient, yoon2016hybrid}. The details of the forward modelling algorithm are given in Appendix A.

\subsection{Inversion}

Most inverse problems are solved using least-squares approaches with some regularizing constraints to improve stability and convergence. The most common formulation is the following regularized nonlinear minimization problem with the quadratic cost functional:
\begin{equation} \label{eq:InvProblem}
\min_{\mathbf{m}} \mathbf{\phi (m)} = \min_{\mathbf{m}} \left( {\left\| \mathbf{F(m) - d} \right\|}_2^2 + \lambda \mathbf{R(m)} \right).
\end{equation}
Here $\mathbf{F(m)}$ is the forward modelling operator, $\mathbf{d}$ is the vector of data observations, $\mathbf{R(m)}$ is the regularization functional, and $\lambda$ is the Lagrange multiplier. The first term on the right-hand side is the $L_2$-norm based data misfit that measures the distance from the observed data to the simulation results. A typical large-scale inverse problem requires optimization in high-dimensional space (millions of model parameters) to minimize the misfit between $\mathbf{F(m)}$ and $\mathbf{d}$ and thus determine the model parameters $\mathbf{m}$. The role of the regularization functional $\mathbf{R(m)}$ is to ensure the well-posedness of the nonlinear inverse problem in the presence of noise and/or inadequate measurements. Typically it penalizes the difference between the current and the reference models ${\left\| \mathbf{m - m_{ref}} \right\|}_\textrm{A}$ in some norm $\textrm{A}$. In practice, a good choice of regularization scheme is often more important than convergence to smaller values of the cost function.

Most local minimization algorithms are iterative, i.e. they start from a starting model $\mathbf{m_0}$ and produce a sequence of approximations $\mathbf{m_k}$ that is expected to converge to the ``true'' model $\mathbf{\hat{m}}$. The model parameters are typically set as the logarithm of resistivity $\mathbf{m}=\ln \mathbf{\rho}$ to decrease variations in the gradient. The model update is written as
\begin{equation} \label{eq:ModelUpdate}
\mathbf{m_{k+1}} = \mathbf{m_k} + \alpha_k \mathbf{p_k},
\end{equation}
where search direction $\mathbf{p_k}$ is typically chosen as a descent direction opposite to the gradient
\begin{equation} \label{eq:UpdatePK}
\mathbf{p_k} = - \mathbf{B_k} \mathbf{g_k} = - \mathbf{B_k} \mathbf{\nabla_{m_k} \phi(m_k)},
\end{equation}
and the gradient of the cost function w.r.t. the model parameters is given by
\begin{equation} \label{eq:Grad}
\mathbf{g} = \nabla_\mathbf{m} \mathbf{\phi (m)} = -\text{Re} \left[ \mathbf{J}^T (\mathbf{F(m) - d})^* \right] + \lambda \nabla_\mathbf{m} \mathbf{R(m)}.
\end{equation}
Using the adjoint-state method, the data gradient can be obtained by solving two forward problems for all sources and frequencies. This constitutes the most computational expensive part for most of deterministic inversion algorithms. Choosing a good value for step length $\alpha_k$ can be challenging as well: small steps lead to slow convergence, while too large values lead to the so-called ``overshooting'' when the cost function fluctuates around the minimum. Building a good search direction $\mathbf{p_k}$ can be expensive since the second-order derivative information is essential. In the steepest descent method, $\mathbf{B_k}$ is simply the identity matrix; in Newton's method, it is the exact Hessian $\nabla^2 \phi(\mathbf{m_k})$, while in quasi-Newton methods $\mathbf{B_k}$ is chosen as some approximation to the Hessian. The NLCG is perhaps the most popular gradient-based method in EM inversion \citep[e.g.,][]{rodi2001nonlinear, commer2008new, moorkamp2011framework}. It requires less computational effort and no additional matrix storage compared to the quasi-Newton and Gauss-Newton schemes, while converging faster than the steepest descent. However, first-order methods are more prone to getting trapped in numerous ``bad'' local minima that are present in all realistic EM inverse problems due to highly non-convex cost functions. In fact, not only local minima but also saddle points (i.e. points where one dimension slopes up and another slopes down) introduce problems to minimization methods. Saddle points are surrounded by a plateau of the same cost function values, which makes it notoriously hard to escape, as the gradient is close to zero in all directions. Second-order optimization methods are more robust but less computationally efficient and have large memory requirements, which limits their practical applications for large-scale datasets.

In practice, local search methods are often able to find an acceptable local minimum (i.e. a model with a sufficiently low value of the cost function) only if a good starting model $\mathbf{m_0}$ is chosen. This dependence on the choice of the starting model is another major shortcoming of traditional inversion. The starting model should contain a priori information about subsurface properties, which is often difficult to estimate reliably for spatially complex geologies. Global optimization methods \citep{sambridge2002monte, sen2013global} overcome some of the above-mentioned issues but they are even more expensive in terms of computational cost. This computational complexity becomes the critical limiting factor in 3-D.

In the next section, I outline a new data-driven approach to inversion based on DL that attempts to avoid the above-mentioned problems. The goal is to identify the subsurface resistivity model directly from the EM data in a single step without constructing the gradients. Instead, a deep NN is trained on multiple synthetic examples to learn the mapping from data space to model space. After training, the network is able to successfully predict the distribution of resistivity from new, previously unseen data.

\section{Deep Learning Inversion}

\subsection{Convolutional neural networks}

Convolutional neural network (CNN) is a specialized kind of networks for processing data that has a grid-like topology \citep{lecun1989backpropagation}. CNNs have played an important role in the history of DL \citep{goodfellow2016deep} and have led to a recent series of breakthroughs in image classification \citep{krizhevsky2012imagenet, simonyan2014very, he2016deep}. They have also been tremendously successful in other practical applications including video analysis and time-series data processing (the latter can be thought of as a 1-D grid taking samples at regular time intervals). CNNs are a key example of a successful application of insights obtained by studying the brain (being to some extent inspired by the structure of the mammalian visual system). As its name indicates, a convolutional network employs a \textit{convolution} operation, i.e. filtering by a feature map or kernel, in place of general matrix multiplication in fully connected networks (in fact, convolution corresponds to a product by a sparse matrix). Deep CNNs with a large number of stacked layers naturally process low and high-level features in data. Almost all CNNs employ an operation called \textit{pooling} to make the representation approximately invariant to small translations of the input. Typically, the \textit{max pooling} operation that outputs the maximum within a rectangular neighbourhood is employed. Pooling with downsampling also reduces the representation size (usually by a factor of two) and hence reduces the computational and statistical burden on the next layer. Pooling is not strictly necessary (downsampling can be also performed directly by convolutional layers with a stride) but is commonly applied in processing rich high-resolution photographs due to its high efficiency.

\begin{figure}
\centering \includegraphics[width=0.7\linewidth]{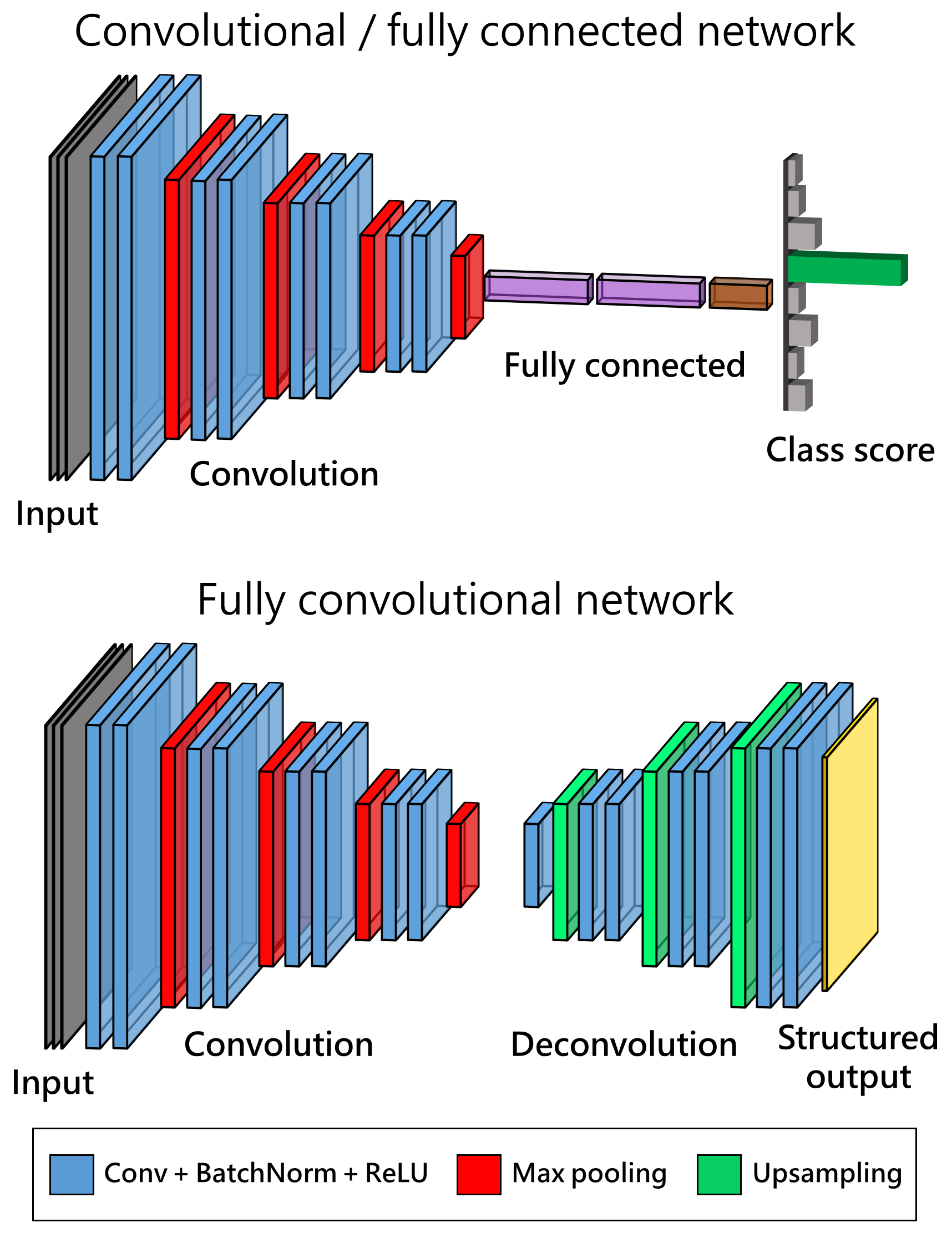}
\caption{Top: conventional convolutional network architecture with fully connected layers at the end, commonly used in classification problems. Bottom: fully convolutional network architecture that outputs a high-dimensional structured object.}
\label{Fig:CNN}
\end{figure}

A typical architecture of a convolutional network is shown in Figure \ref{Fig:CNN}a. The network consists of a set of layers each of which contains several filters for detecting various features in the input. The most common CNN architectures have been developed for classification (predicting a class label) and regression (outputting a real value) tasks and have a fully-connected layer at the end \citep[e.g.,][]{krizhevsky2012imagenet, simonyan2014very, he2016deep}. At the same time, CNNs can also be used to output a high-dimensional structured object. These networks are often based on the fully convolutional architecture (Figure \ref{Fig:CNN}b), the most notable examples being the image segmentation networks \citep{long2015fully, badrinarayanan2017segnet}. Geophysical inversion is an example of a structured output task since its result, the subsurface model, is a structured 2-D or 3-D object (spatial output maps). This makes the fully convolutional architecture a natural choice for inversion. Discarding the fully connected layers also significantly reduces the number of parameters in the network and makes it easier to train.

In common fully convolutional network applications such as image segmentation and video analysis, the size of the output is equal to the size of the input. In inversion, however, there is no general relationship between the dimensions of the data space and model parameter space. This does not pose a serious difficulty, since there are many ways to obtain an object of size $m \times n$ from $a \times b$ data, e.g., by a combination of max pooling layers, upscaling layers, cropping operations, and padding. In the following examples, I use a combination of max pooling layers with stride two followed by a combination of upscaling layers and cropping at the end. This architecture is similar up to some extent to the convolutional encoder-decoder \citep{badrinarayanan2017segnet}. The inner convolutional layers use the ``same'' padding to keep the size of their output equal to the size of the input. This prevents the representation from shrinking with depth and allows to create an arbitrarily deep network.

CNNs naturally use multichannel convolution on the input. Each channel is an observation of a different quantity at some point in space (e.g., color channels in image data). For EM measurements, possible choices of the channels are the electric and magnetic field components, amplitude and phase, frequencies. CNN models can efficiently scale to very large sizes and can handle inputs of varying size (e.g., different numbers of electric and magnetic receivers). When the receivers are located in a grid-like fashion, I employ a classical 2-D convolutional network with the input layer being a 3-D array of size $w \times h \times d$, where $w$ and $h$ are the spatial dimensions of the received grid, and $d$ is the number of channels. This is similar to image processing tasks with input images of $w \times h$ pixels and $d$ color channels. For other acquisition geometries, I put the measurements into a 2-D $l \times d$ array and employ a 1-D CNN.

A rectified linear unit (ReLU) and its variants \citep{xu2015empirical} are perhaps the most used \textit{activation functions} for CNNs as well as other types of networks. The main benefits of ReLUs are sparsity, reduced likelihood of vanishing gradient, better convergence performance, and cheap computational cost. The so-called dying ReLU problem can be handled by using Leaky ReLU instead of standard rectified linear units. Nowadays, ReLUs almost replaced the standard sigmoid activation functions in hidden layers of the network (sigmoids are still widely used in the output layer for classification). Another common operation that often has a dramatic effect on CNN performance is \textit{batch normalization}. It allows for much faster training, higher accuracy, and has low sensitivity to initialization parameters. Besides that, batch normalization is useful for regularization purposes.

\subsection{Regularization}

Successful training of the network will result in a small error on the \textit{training dataset}, called the training error. In practice, we are interested in how well the algorithm performs on data that it has not seen before since this determines its real-world performance. This ability is called \textit{generalization}. All performance metrics must be calculated on a \textit{test dataset} that is separate from the data used for training (though it often comes from the same distribution). In other words, we want the test error, also called the generalization error, to be low as well. Two key factors determining the performance of a DL model are its ability to (a) achieve low training errors and (b) make the gap between training and test errors small. Large training errors correspond to \textit{underfitting}, which is usually easy to detect and fix by changing the model capacity or training algorithm. A large gap between the training and test errors corresponds to \textit{overfitting}. There are several ways to reduce overfitting for a selected model, including changing the regularization scheme and reducing architecture complexity (thus decreasing the number of trainable parameters). In many cases, overfitting happens due to the training dataset being small; thus, data augmentation might help (Section \ref{sec:data}).

Regularization aims at reducing the generalization error of a learning algorithm, often by sacrificing the training error. There is no best approach to regularization but a wide range of tasks typically can be solved effectively using general-purpose forms of regularization. \textit{Dropout} and \textit{early stopping} are probably the simplest and the most commonly used regularization techniques in DL applications. Dropout \citep{srivastava2014dropout} drops out units of a network (both hidden and visible) avoiding training all nodes on all training data, thus decreasing overfitting. Dropout is very computationally cheap and works well with nearly any model that uses a distributed representation \citep{goodfellow2016deep}. On the other hand, dropout, as well as other regularization techniques, typically require increasing the network size. Thus, lower test set errors achieved using dropout technique come at the cost of larger networks and hence more iterations of the training algorithm. Dropout is also generally less effective for CNNs. The recent trend in modern convolutional architectures is to use dropout only on fully-connected layers and use batch normalization between convolutions \citep{ioffe2015batch, he2016deep} to regularize the model and make its performance more stable during training.

DL models have \textit{hyperparameters}, i.e. settings that one can use to control the desired behaviour. The values of hyperparameters are not adapted by the learning algorithm itself; it is not appropriate to learn the hyperparameters that control model capacity on the training set (which would lead to overfitting). Instead, a third dataset called \textit{validation dataset} is made of examples that the training algorithm does not observe and used for hyperparameters tuning. Validation dataset can also be used for regularization by early stopping, i.e. stopping the training when the errors on this set increase, as the network starts overfitting to the training data. The idea of early stopping is that rather than choosing the model which performs better on the training set (and possibly overfits the data), we choose an earlier model with better validation set error (and thus, hopefully, better test set error). In the proposed scheme, I use both dropout and early stopping together. The proportions for the training, validation, and test datasets are discussed in Section \ref{sec:data}.

\subsection{Loss functions}

Training a NN is an optimization problem equivalent to finding the minima of the \textit{loss function}, which measures how ``good'' is the predicted model. Evaluating the cost function \eqref{eq:InvProblem} for each model would be prohibitively expensive and not necessary since true models are available during the training stage. Thus, instead of comparing the EM response between a given model and the observed data, we estimate the difference between the two models (similar to the regularization term $\mathbf{R(m)}$ in deterministic inversion). Choosing a proper loss function that leads to the desired behaviour of the method is a nontrivial task. Two common metrics to measure accuracy for continuous variables without considering their direction are the Mean Absolute Error (MAE) and Root Mean Squared Error (RMSE). The MAE is the average of the absolute values of the differences between forecast and the corresponding observation:
\begin{equation} \label{eq:MAE}
\textrm{MAE} = \frac{1}{n} \sum_{i=1}^{n}{|m_i - \hat{m}_i|}.
\end{equation}
The MAE is a difference in $L_1$ norm, i.e. a linear score where all the individual differences are weighted equally in the average. The RMSE is a quadratic scoring rule (difference in $L_2$ norm) that squares the errors before they are averaged:
\begin{equation} \label{eq:RMSE}
\textrm{RMSE} = \sqrt{\frac{1}{n} \sum_{i=1}^{n}{(m_i - \hat{m}_i)^2}}.
\end{equation}
The RMSE gives a relatively high weight to large errors, hence it is most useful when large errors are particularly undesirable. The RMSE will always be larger or equal to the MAE. Both these metrics can be used together to monitor the variation in the errors. The greater the difference between them, the greater the variance in the individual errors. In the following examples, the model parameters are chosen as the logarithm of resistivity, thus decreasing variations in \eqref{eq:MAE} and \eqref{eq:RMSE}.

Another evaluation metric widely used in computer vision for comparing the similarity of two sample sets is the Intersection over Union (IoU, also known as the Jaccard similarity coefficient). IoU is defined as the size of the intersection divided by the size of the union of the sample sets:
\begin{equation} \label{eq:IOU}
\textrm{IoU} = \frac{\lvert A \cap B \rvert}{\lvert A \cup B \rvert} = \frac{\lvert A \cap B \rvert}{\lvert A \rvert + \lvert B \rvert - \lvert A \cap B \rvert}.
\end{equation}
The IoU loss function is defined as $1 - \textrm{IoU}$ and measures dissimilarity between sets (also known as the Jaccard distance). The main limitation of IoU is that it is defined in terms of binary set membership. Though some extensions do exist for multisets, I use this measure as the training loss function only in those examples where a target/non-target mask is predicted rather than actual resistivity values. Given the diffusive nature of EM fields, IoUs of 0.4--0.5 would be normally considered as sufficiently accurate, while values around 0.6--0.7 or above indicate a very good match (see also the Discussion section).

\subsection{Training}

Most CNNs are trained in a classical supervised fashion using full forward and back propagation through the entire network. The weights are assigned during the training stage using an optimization algorithm that minimizes the loss function. Standard (batch) gradient descent optimization methods update the model after the complete computation of the gradient for the entire training dataset, which becomes computationally expensive for large datasets. Variants of the stochastic gradient descent (SGD) algorithm are the dominant training methods for DL models nowadays. The idea behind this algorithm is that the gradient is an expectation, which may be approximately estimated using a small set of samples. The SGD and mini-batch algorithms allow updating the parameters on each iteration using only one sample or a mini-batch of samples, respectively. Mini-batch methods have more stable convergence compared to the SGD and are commonly applied in deep NN training nowadays.

In the following examples, I use the Adam optimization algorithm \citep{kingma2014adam}, which is a rather common choice due to its computational efficiency and fast convergence. In Appendix B, I also compare the performance of all standard optimization algorithms, namely Adagrad, Adadelta, RMSprop, Adam, AdaMax, and Nadam. Training very deep networks is computationally intensive and requires large computing power. Graphics processing units (GPUs) were originally developed for graphics applications; next, their performance characteristics turned out to be beneficial for scientific computing and later for deep NNs. GPUs have a high degree of parallelism and high memory bandwidth (at the cost of having a lower clock speed compared to CPUs) and clearly outperform CPUs in network training. A recent alternative to GPUs is Google's tensor processing units (TPUs) developed specifically for DL applications.

\subsection{Data preparation}
\label{sec:data}

The amount of data required to achieve acceptable performance depends on a particular application. A rough rule of thumb is that a supervised DL algorithm will generally work well with around 5,000 labeled examples per category, though smaller datasets can be used in problems where data generation is expensive \citep{goodfellow2016deep}. Test set performance being much worse than training set performance (overfitting) often indicates a lack of data. However, obtaining more data is not always possible. In these cases, \textit{data augmentation} can significantly improve performance. Dropout can also be seen as a process of constructing new inputs by multiplying by noise (thus playing a role similar to data augmentation).

The most common method to reduce overfitting on image data is to artificially enlarge the dataset. While it is relatively easy to perform data augmentation in image processing with very little computation (e.g., randomly rotating the image, adding reflections, zooming in, adding a color filter, etc.), it is often non trivial to do so in modelling problems. In Appendix A, I describe the methods for faster simulations of many similar forward problems and the data augmentation using model symmetry, which allows for a fourfold increase of the simulated dataset.

There are different ways to split all available data into the training, validation and test sets. Commonly occurring ratios are 70/15/15 and 80/10/10, though in the modern big data era, datasets of millions and tens of millions of examples are often split as 98/1/1 or even 99/0.5/0.5. These proportions also depend on the training methodology employed.

\subsection{Network specifications}

The optimal network architecture for a particular task and best hyperparameters must be found by trying different models and monitoring the training and validation errors. In this study, based on an extensive series of experiments, I use the fully convolutional architecture (Figure \ref{Fig:CNN}b). Two remaining architectural considerations are choosing (a) depth of the network and (b) number of convolutions and their width at each network level. Deep networks achieve low training errors very fast but they overfit the data and generalise poorly for new samples. Very deep networks often face the so-called degradation problem when accuracy rapidly degrades with the depth increasing; this issue can be addressed, for example, by residual learning \citep{he2016deep}.

\begin{figure}
\centering \includegraphics[width=1.0\linewidth]{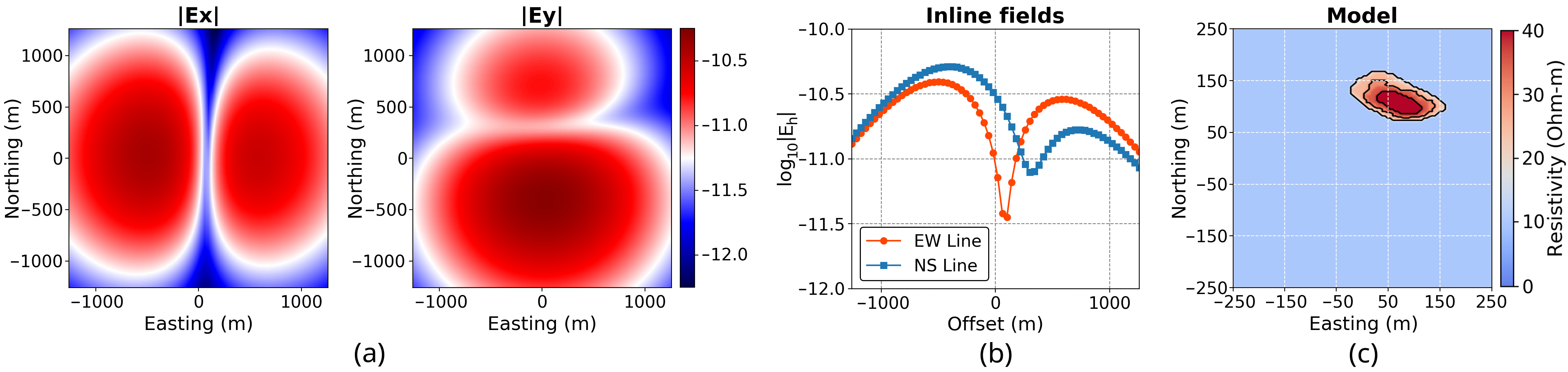}
\caption{Difference in the amplitude of the horizontal electric field measured on the grid of surface receivers (a); difference in the amplitude of the inline component of the electric field at two lines of receivers (b); resistivity model as the network output (c).}
\label{Fig:DataToModel}
\end{figure}

The 2-D CNNs employed in the following examples consist of two sets of convolutional blocks with max pooling layers followed by three sets of convolutional blocks with upscaling layers. There are no fully connected layers in the networks, which makes them belong to the class of fully convolutional networks. Each convolutional block consists of a sequence of convolutions with $3 \times 3$ filters, batch normalization, and leaky ReLU activation. Max pooling is performed over a $2 \times 2$ pixel window, with stride 2. The number of filters starts from 32-64 in the first layer (depending on the number of field components in the input) and then increases by a factor of 2 after each max pooling layer. After each upscaling layer, I decrease the number of filters by the same factor. The output layer enables pixel-wise prediction of the resistivity model. Figure \ref{Fig:DataToModel} shows the examples of network's input and output, namely the difference in the amplitudes of the horizontal electric field between the monitor and baseline data, and the resistivity model. Resistivity changes at the reservoir depths caused by $\text{CO}_\text{2}$ movement lead to observable differences in the EM responses at the surface.

A batch normalization layer is used after each convolutional layer to avoid problems with different scales of the parameters. Based on a series of tests, I choose the leaky ReLU with parameter 0.1 as the activation function, which has been found the most efficient option among other alternatives. Even though batch normalization has a regularizing effect and for some applications it has completely replaced dropout, I employ the latter as well due to relatively small datasets used in training. Dropout layers with a rather small rate of 0.1 are applied after each convolutional block to reduce overfitting.

For implementation, I use open-source libraries TensorFlow \citep{abadi2016tensorflow} and Keras \citep{chollet2015keras} that support all of the above-mentioned DL concepts and optimization algorithms. The data is stored in a single-precision float32 format for efficiency. Using half precision float16 format on modern accelerators can reduce the computational time and memory demand even further though it also impacts the training accuracy. Training of the networks shown in Section 4 was performed on a system equipped with NVIDIA P100 Pascal GPU.

\section{Numerical examples}

\begin{figure}
\centering
 \subfigure{\includegraphics[width=0.41\textwidth]{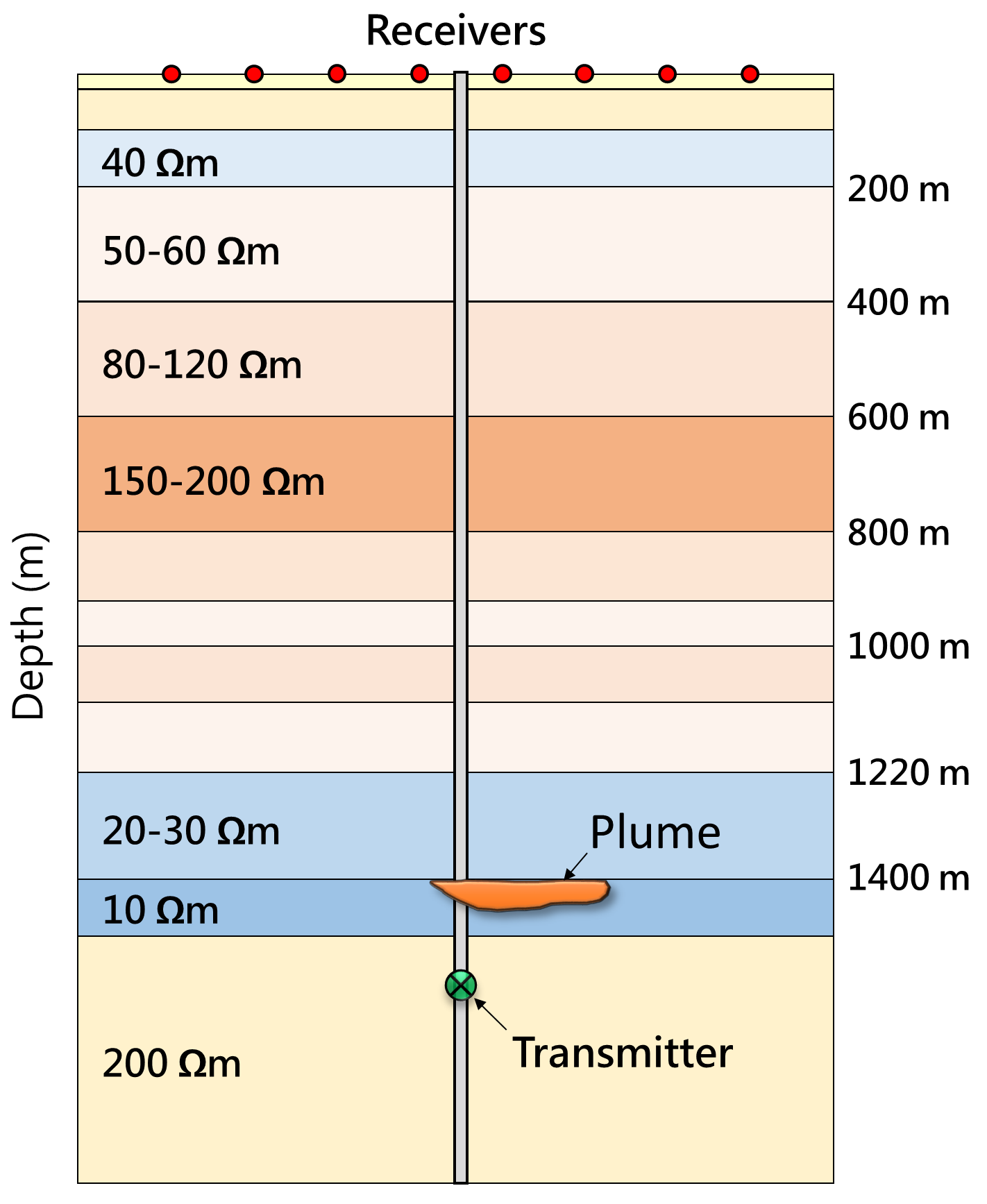}}
 \hspace{0.5cm}
 \subfigure{\includegraphics[width=0.475\textwidth]{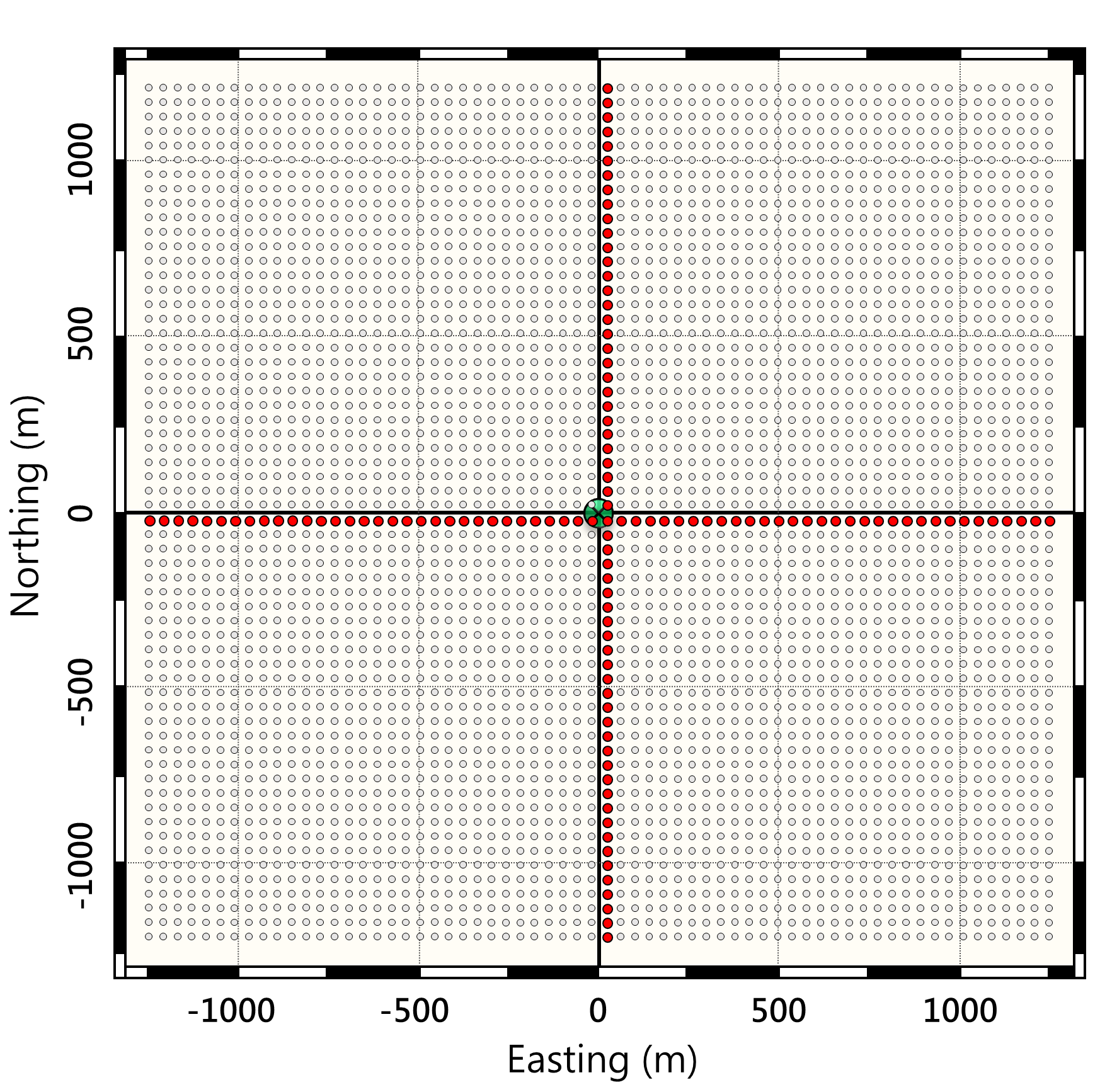}}
\caption{Left: background resistivity model and setup of the borehole-to-surface CSEM. Right: grid of receivers at the surface. In the idealized case, the input data includes the measurements at all $64 \times 64$ receivers. In the practical case, only the measurements at two lines of receivers shown in red are used in inversion.}
\label{Fig:SurveySetup}
\end{figure}

To illustrate the performance and applicability of the proposed inversion scheme, I show several numerical examples that arise from an onshore CSEM $\text{CO}_\text{2}$ monitoring scenario. Monitoring $\text{CO}_\text{2}$ plume movement and extent is a critical component of carbon capture and storage projects. Downhole EM methods are sensitive to changes in phase saturation, which makes them well-suited for $\text{CO}_\text{2}$ monitoring tasks. The CNNs discussed in the previous section are trained on synthetic frequency-domain EM data simulated for various resistivity models in an onshore CSEM monitoring scenario. The baseline subsurface model (Figure \ref{Fig:SurveySetup}a) is the same in all examples and follows the 18-layer model of the Hontomin $\text{CO}_\text{2}$ storage site used in \cite{puzyrev2017three}. The layer thicknesses vary from 20 to 80~m; the primary reservoir is located at a depth of 1400~m. This is the most conductive unit of the model (10 Ohm-m), and it contains the injected $\text{CO}_\text{2}$. The source is a vertical electric dipole located at $z=1500 m$ in the well; electric and magnetic field receivers are located at the surface. This borehole-to-surface configuration is known to have good sensitivity to resistors located in the vicinity of the well. Three frequencies of 8, 32, and 128 Hz are used in inversion.

$\text{CO}_\text{2}$ is sequestrated in a saline aquifer characterized by highly conductive brine and low clay content. I use Archie's equation to estimate the reservoir's resistivity change due to the $\text{CO}_\text{2}$ injection. Assuming the presence of two fluid phases (formation brine water and injected $\text{CO}_\text{2}$), the post-injection resistivity ${\rho_1}$ is related to the resistivity of formation saturated only with brine water ${\rho_0}$ as
\begin{equation} \label{eq:Saturation}
\frac{\rho_1}{\rho_0} = \left( 1 - S_{CO_2} \right)^{-n}.
\end{equation}
Assuming the saturation exponent $n=2$, a homogeneous saturation of $S_{CO_2}=0.5$ results in a rather small resistivity contrast of 4. This is the maximum contrast between the anomalies and surrounding formations in the following examples. Considering the reservoir geometry, pressure and temperature conditions, the thickness of the simulated $\text{CO}_\text{2}$ plumes is chosen as 40 m. Under the assumption that the plumes do not penetrate the seal formation, the unknowns are their shapes and positions in the horizontal plane. The $500 \times 500$ m zone of interest of our pixel-based inversion is discretized in $x$ and $y$ directions by $5 \times 5$ m square blocks (10,000 blocks in total) and the algorithm determines the resistivity of each block separately.

\begin{figure}
\centering \includegraphics[width=1.0\linewidth]{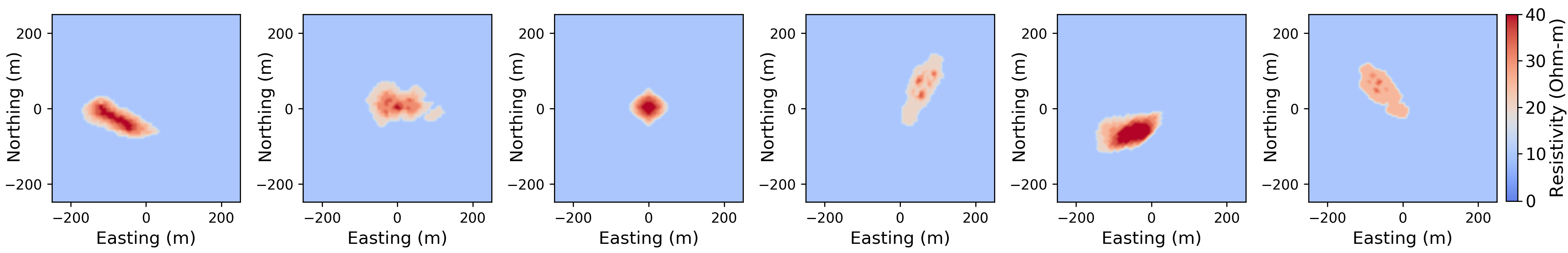}
\caption{Representative set of resistivity models used in network training. $\text{CO}_\text{2}$ plume shapes range from geometric primitives such as circles and ellipses to realistic plumes estimated at Sleipner and Ketzin storage sites.}
\label{Fig:PlumesSamples}
\end{figure}

In Figure \ref{Fig:PlumesSamples}, I show a representative set of plumes (resistivity models in the horizontal plane) used for network training. Plume shapes range from primitives to realistic plumes estimated at offshore $\text{CO}_\text{2}$ storage site at Sleipner, Norwegian North Sea \citep{chadwick2010quantitative}, and onshore site at Ketzin, Germany \citep{martens2012europe}. Their lateral dimensions vary from 40~m to more than 200~m. In the first two examples shown below, the $\text{CO}_\text{2}$ saturation is assumed homogeneous in the entire plume and thus the anomalies have constant resistivity of 40 Ohm-m. In the third example, the saturation varies from 0.1 to 0.5 and the plumes have variable resistivity as shown in Figure \ref{Fig:PlumesSamples}.

\subsection{Plume delineation}

\subsubsection{Example 1: Receiver grid}
\label{sec:Ex1}

In the first example, the EM field is recorded on a grid of surface receivers. This setup is quite common in marine CSEM exploration and 3-D seismic surveys and allows to illustrate the connection to image processing CNNs. The network input has a fixed size of $64 \times 64$ according to the number of surface receivers employed. Such a high number of perfectly-aligned receivers serves demonstration purposes only and is not necessary in practice, as will be shown in the following example. Missing or noisy measurements can be safely excluded from the input image or interpolated from the neighbouring locations without significant accuracy loss (cf. image reconstruction and super-resolution tasks using DL). The input data has 10 channels for each frequency, namely the logarithm of the amplitude and phase of the $E_x$, $E_y$, $H_x$, $H_y$, and $H_z$ field components (as previously shown in Figure \ref{Fig:DataToModel}a). Using the three frequencies mentioned above, the network receives a $64 \times 64 \times 30$ array of floats as input. Using fewer input channels, e.g., only amplitudes and phases of the horizontal electric field components, also allows for successful resistivity prediction, though of slightly lower quality and reliability.

The full dataset used in this example consists of 20,000 different resistivity models and their corresponding EM responses. 5,000 of them were generated using full 3-D simulations of models containing plumes of various sizes, shapes, and positions. In this example, the saturation is assumed constant and equal to 0.5, thus resulting in a resistivity contrast of 4 between the plume and the host formation. The rest 15,000 examples were obtained using data augmentation (Appendix A). The training dataset consists of 18,400 examples, while the validation and test sets have 800 models each, which corresponds to a 92/4/4 split. The data used in network training is noise-free. After training, I apply the network to predict models from the test dataset containing both noise-free and noisy data. These examples were not given to the network during training. For each task, networks with different numbers of convolutional blocks and filters in each convolution were tested and the one which performed best on the validation dataset was chosen.

\begin{figure}
\centering \includegraphics[width=1.0\linewidth]{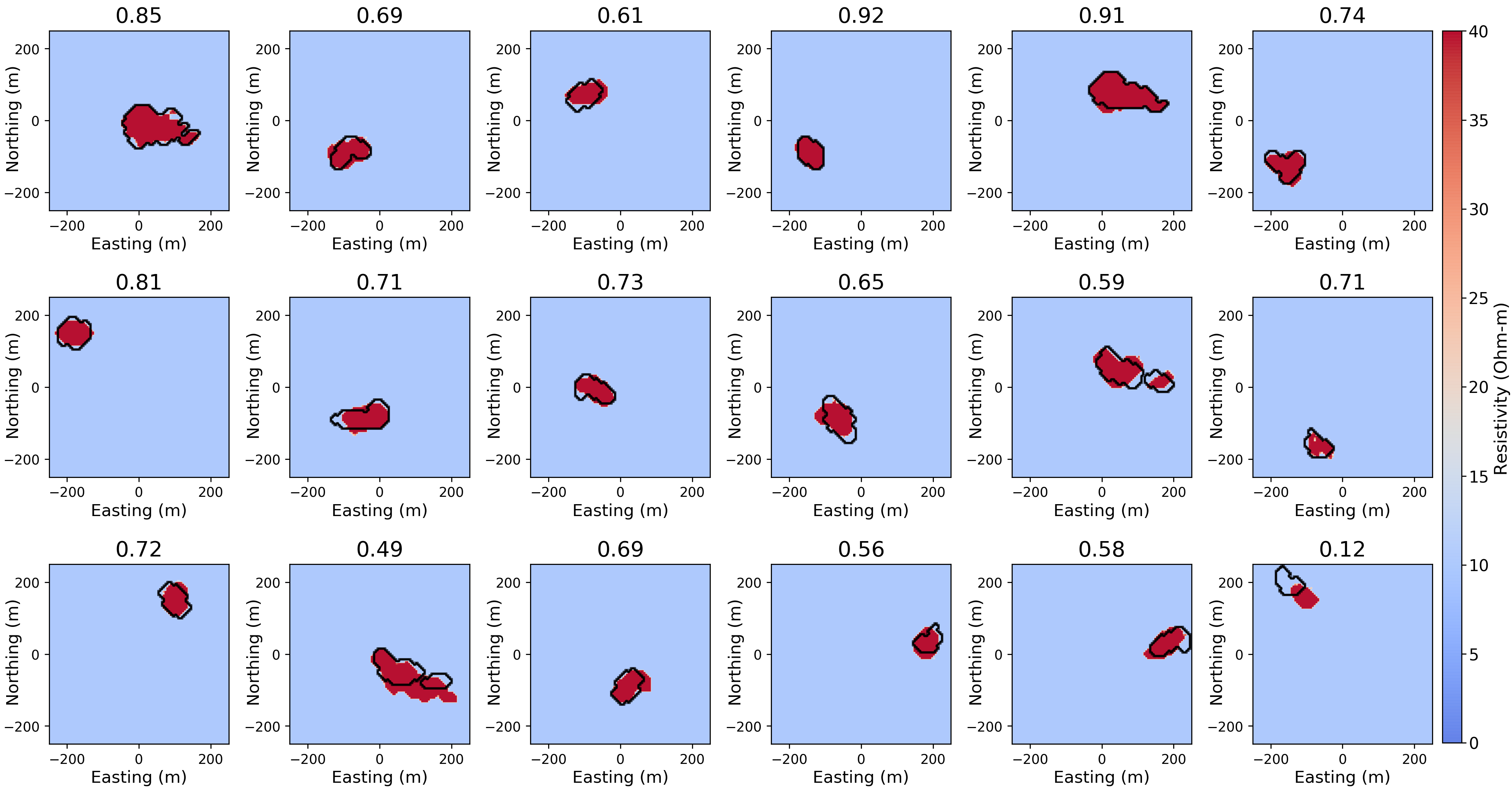}
\caption{Resistivity models predicted by the 2-D CNN versus the true models (black contours). The IoU value for each model is shown in the image title. The average IoU of the test dataset is 0.67.}
\label{Fig:Results1_2D_zeronoise}
\end{figure}

Figure \ref{Fig:Results1_2D_zeronoise} compares the predicted and true resistivity models. Both sets have a high degree of similarity: 18\% of the predictions are excellent (IoU $\ge 0.8$) and other 54\% are very good ($ 0.6 \le$ IoU $ < 0.8$). The average IoU of the entire test dataset is 0.67. The network is not restricted to binary output; its last layer outputs a real number (logarithm of resistivity) for each cell of the model. However, as the training set in this example contains only two values of resistivity (10 Ohm-m background and 40 Ohm-m plume), the network learns this format after the first few epochs of training. In Section \ref{sec:Ex3}, I show a similar network trained on models with varying resistivity.

\begin{figure}
\centering \includegraphics[width=1.0\linewidth]{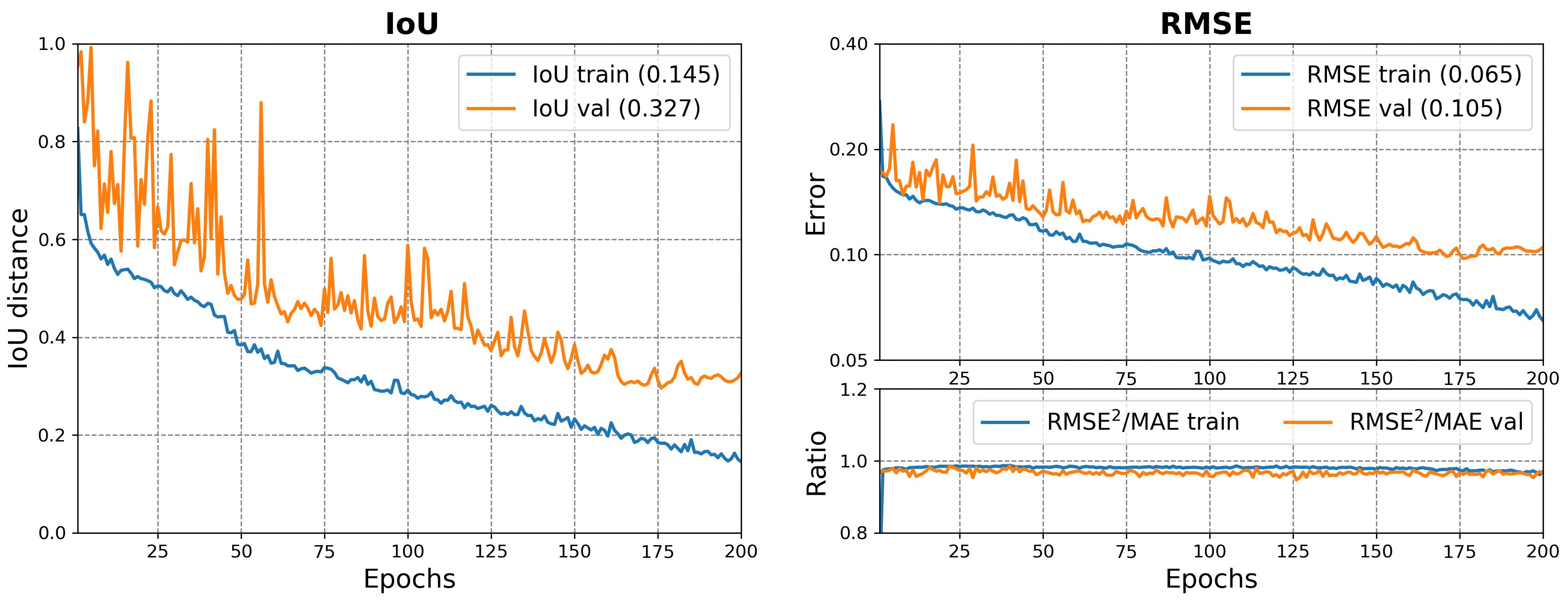}
\caption{The training and validation losses for a hyperparameter optimized 2-D CNN using Adam algorithm for optimization. Both IoU (left) and RMSE (right) validation losses start growing around epoch 180, indicating overfitting. The RMSE metric is normalized such that the resistivity values corresponding to zero and 50\% saturation are mapped to $[0, 1]$ interval. The MAE metric is omitted here since it highly correlates with the RMSE (bottom right), which is expected for a network that produces a nearly binary output.}
\label{Fig:TrainingHistory}
\end{figure}

The hyperparameters of the network used in this example were tuned to minimize the validation errors. The network consists of five levels (two max pooling and three upscaling) each having three convolutional blocks (convolution + batch normalization + leaky ReLU). Using 64-256 filters with dropout at each layer achieves better results compared to smaller networks without dropout, despite the larger training errors. The network has 2.7 million trainable parameters, which is not large by modern standards and does not require extensive computational resources for training. Figure \ref{Fig:TrainingHistory} shows the training history over 200 epochs using the Adam optimizer with batch size of 32 and the default learning rate of 0.001. We observe that the training error monotonically decreases, while the validation error fluctuates but decreases as well. In total, the normalized RMSE of the training set decreases 5 times during training. The overfitting starts around epoch 180. For such relatively small datasets, dropout is essential as it helps to prevent overfitting, thus leading to significantly better generalization. 

\begin{figure}
\centering \includegraphics[width=1.0\linewidth]{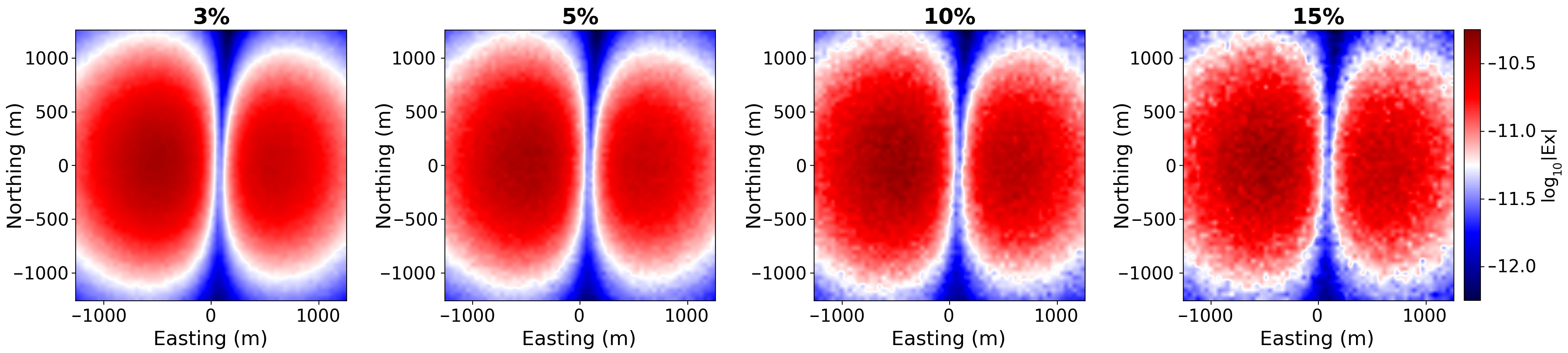}
\caption{Network's input (amplitudes of the horizontal electric field) at different noise levels (normally distributed random noise with relative variance of 3\%, 5\%, 10\%, and 15\%). Noise-free measurements for the same example were shown previously in Figure \ref{Fig:DataToModel}a.}
\label{Fig:NoisyInput}
\end{figure}

Next, in order to test the method in a realistic setting, I apply it to noisy datasets. Normally distributed random noise with zero mean and specified variance is added to both the validation and test datasets. Figure \ref{Fig:NoisyInput} shows the effect of different noise levels on the network's input. The average IoUs over these noisy test datasets are given in Table \ref{Tab:Results3_NoiseExample}. The IoUs of the predicted and true resistivity models for the measurements with 3-5\% noise are very similar to the noise-free case. As the noise level increases, the number of bad predictions (IoU $ < 0.2$; often IoU $= 0$) rapidly grows though the average IoU remains quite high. When the noise exceeds certain threshold (typically around 23-25\%), the network stops making reliable predictions and its IoU drops to nearly zero.

\begin{table}
\begin{center}
\begin{tabular}{ |c|c|c| } 
 \hline
 \textbf{Noise level} & \textbf{Average IoU} & \textbf{Bad predictions} \\
 & & \textbf{(IoU $ < 0.2$)} \\
 \hline
 \textit{0} & 0.672 & 17 \\ 
 \hline
 \textit{3\%} & 0.663 & 18 \\ 
 \hline
 \textit{5\%} & 0.649 & 23 \\ 
 \hline
 \textit{10\%} & 0.625 & 51 \\ 
 \hline
 \textit{15\%} & 0.581 & 103 \\ 
 \hline
\end{tabular}
\caption{Network accuracy for different levels of noise in the test data. The test dataset consists of 800 examples.}
\label{Tab:Results3_NoiseExample}
\end{center}
\end{table}

\begin{figure}
\centering \includegraphics[width=1.0\linewidth]{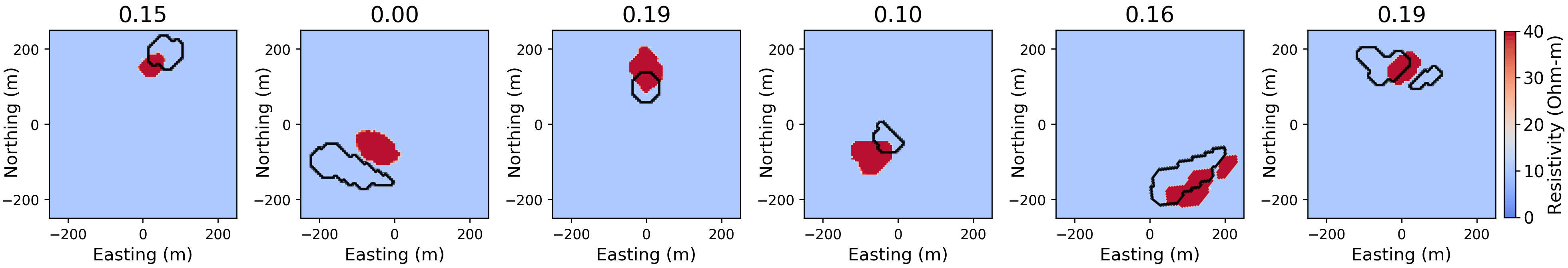}
\caption{Badly predicted resistivity models (IoU $ < 0.2$) and their true shapes (black contours).}
\label{Fig:Results2_2D_bad}
\end{figure}

Even for the noise-free test data, there still remains a small subset of models where the prediction quality is low (IoU $ < 0.2$), however, their number is around 2\%. Figure \ref{Fig:Results2_2D_bad} shows some of these unsuccessful predictions whose total number is 17 out of 800 models tested. Visual inspection of these images reveals that most of them have resistivity structure that caused the network to make an incorrect prediction. In practice, the accuracy of the method can be controlled by calculating the data misfit term in Equation \eqref{eq:InvProblem}, which can be performed at a cost of one forward modelling. In those rare cases when the DL inversion fails, a deterministic inversion algorithm such as the NLCG or Gauss-Newton can be employed using the network prediction as a starting model.

\subsubsection{Example 2: Receiver lines}
\label{sec:Ex2}

CNN-based approaches have been the most successful on a two-dimensional image topology but they are not restricted to it. The fully convolutional architecture presented here can be easily extended to inputs of other dimensions as well. In the next example, I consider as input data only the measurements at two perpendicular lines of receivers shown in red in Figure \ref{Fig:SurveySetup}b.

There are several ways to modify the network architecture to handle new inputs. One is to continue using 2-D inputs and reshape the input data into a $2 \times 64$ array for each channel. However, in order to illustrate the flexibility of the method, I use another approach which takes $1 \times 128$ arrays as input. The first layers of the network employ 1-D convolution and max pooling operations. Next, after a few 1-D upscaling operations, the structures at hidden layers are reshaped into 2-D arrays and the remaining layers of the network are similar to the previous example. Several architectures were compared based on the validation set errors. The network used in the following example has 0.6 million parameters, which is considerably less compared to the fully 2-D convolutional networks of similar depth used in the previous example. The training is stopped after 250 epochs when the validation error starts to grow. The best IoU achieved for the validation and test datasets is 0.65 and 0.61, respectively.

\begin{figure}
\centering \includegraphics[width=1.0\linewidth]{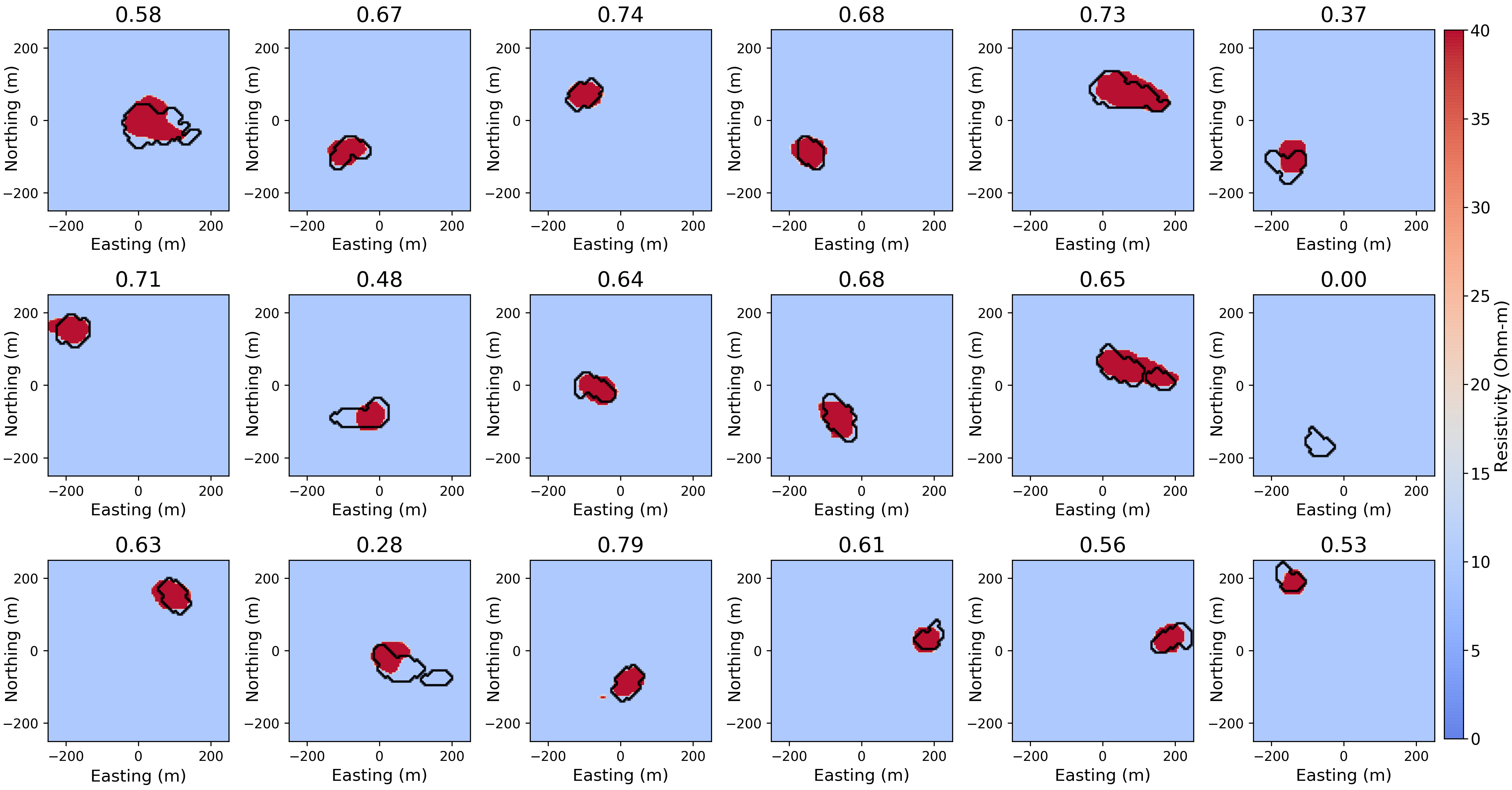}
\caption{Resistivity models predicted by the 1-D CNN versus the true models (black contours). The IoU value for each model is shown in the image title. The average IoU of the test dataset is 0.61.}
\label{Fig:Results4_1D}
\end{figure}

Figure \ref{Fig:Results4_1D} compares the resistivity models predicted by the 1-D network with the true models. Despite the limited data availability in this example, the prediction quality is only slightly worse compared to the previous case. 64\% of the predictions are very good or excellent (IoU $\ge 0.6$). The boundaries of the anomalies are less reliably predicted compared to their positions. The number of badly-predicted models (IoU $ < 0.2$) doubles compared to the previous case and is equal to 39 out of 800.

\subsection{Inversion for resistivity values: Example 3}
\label{sec:Ex3}

In the last example, I consider another dataset of 20,000 models that have varying $\text{CO}_\text{2}$ saturation (and hence electrical resistivity) inside the plumes. The saturation varies from 0.1 to 0.5 and the corresponding resistivities are obtained using Equation \eqref{eq:Saturation}. The network architecture is similar to the one described in Section \ref{sec:Ex1} and takes a $64 \times 64 \times 30$ input.

\begin{figure}
\centering \includegraphics[width=1.0\linewidth]{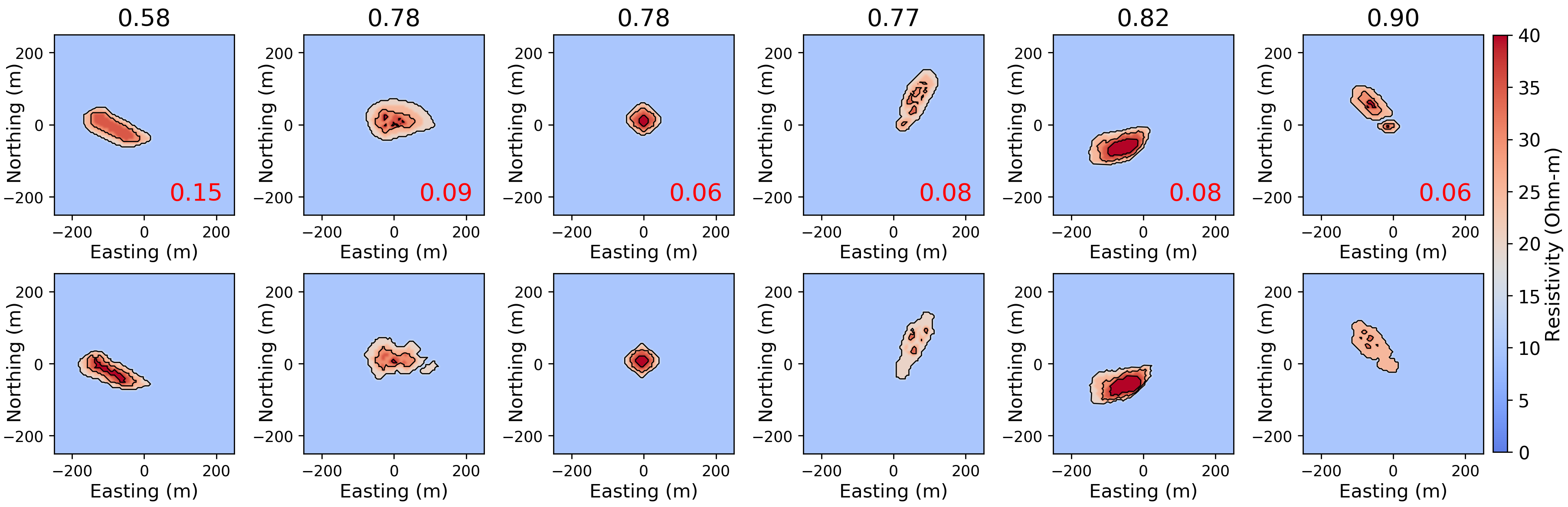}
\caption{Resistivity models predicted by the 2-D CNN (top row) versus the true models (bottom). The average IoU (20 Ohm-m threshold) of the test dataset is 0.65.}
\label{Fig:Results5_2D_Inversion}
\end{figure}

\begin{figure}
\centering \includegraphics[width=1.0\linewidth]{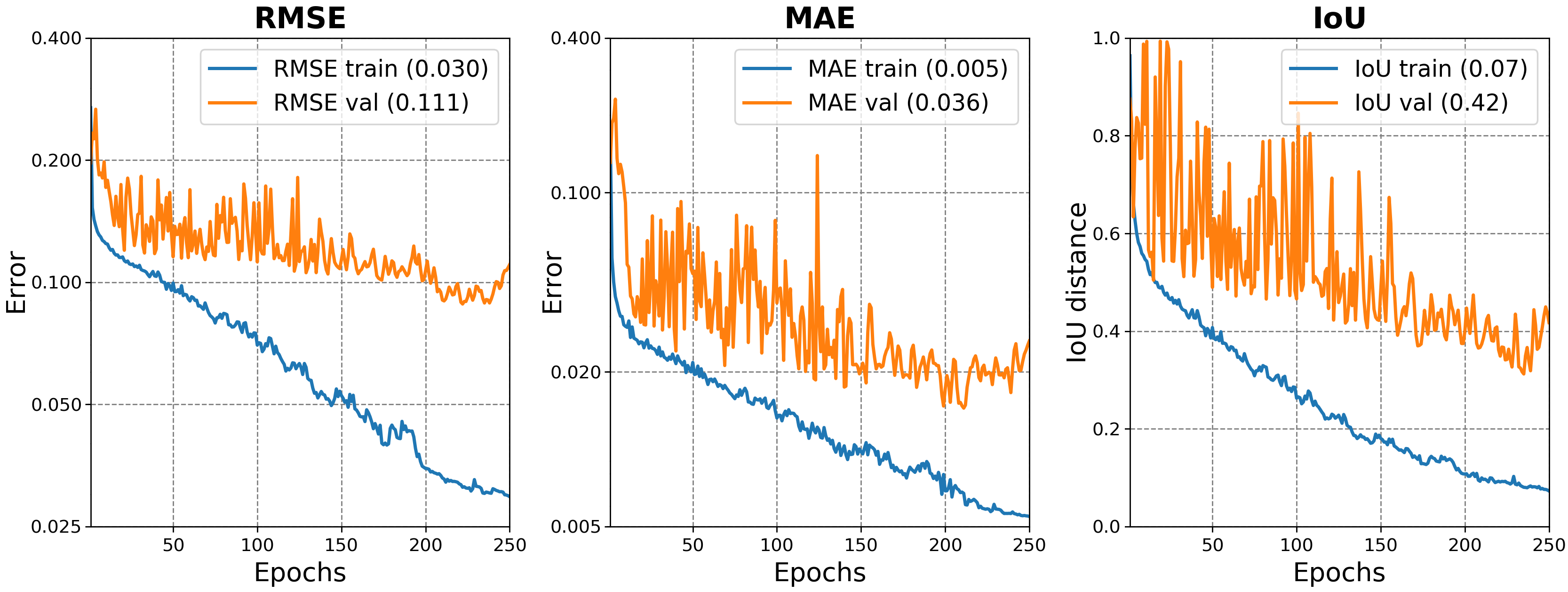}
\caption{The training and validation losses for a hyperparameter optimized 2-D CNN using Adam algorithm for optimization.}
\label{Fig:TrainingHistory2}
\end{figure}

Figure \ref{Fig:Results5_2D_Inversion} compares the resistivity models predicted by the network with the true models. Once again, the CNN predictions are quite similar to the true resistivity distributions. The network can reliable estimate the shape and position of the anomaly; the predicted resistivity is typically lower than the actual values, indicating that the network tends to underestimate it. The average IoU (calculated using a 20 Ohm-m threshold) of the test dataset is 0.65. Figure \ref{Fig:TrainingHistory2} shows the training history. The network used in this example has more filters compared to those used in the two previous sections and hence it requires more epochs to achieve minimum errors on the validation dataset. RMSE \eqref{eq:RMSE} is used as the training loss function.

The number of trainable parameters in the fully convolutional networks used in these three examples ranges between 0.6 and 5 millions. Using deeper networks has little effect on this relatively small dataset. CNNs with slightly different number of filters and dropout rates provide resistivity distributions that also have small differences, hence allowing for some uncertainty quantification by using multiple networks on the same dataset.

\section{Discussion}

One of the main practical limitations of traditional deterministic inversion is its high computational cost caused by a large number of forward modelling operations for multiple source positions and frequencies. Probabilistic methods with all their strengths are even more computationally demanding. This large computational cost often constrains the usage of inversion-based interpretations in 2-D and 3-D. Additional factor that complicates the interpretation of EM data is high level of noise. All this motivates the development and practical use of new inversion methods that have modest computational demands, while being robust and efficient. Now we stand on the threshold of a new era in computational geosciences when DL methods can take simulations of geophysical processes to a new level.

This work has shown the potential of DL methods in EM inversion. To my knowledge, this is the first application of deep CNNs to EM inverse problems. The method does not require the regularization in its common-sense meaning. Instead, the network is trained on a dataset containing realistic models and thus learns how to reproduce a similar model that fits well the data. The main advantage of the DL-based inversion is its high computational efficiency. The networks used in this study have a modest number of trainable parameters (ranging from 0.6 to 5 millions). Using NVIDIA Tesla P100 GPU, training a single network takes between 1.5 and 6 hours depending on its size. Once trained, the network can predict resistivity models from new data in less than one second using a laptop CPU or GPU. Furthermore, deep networks can capture hidden dependencies in data.

Deep learning inversion consists of three stages: data generation, model training and model prediction from a given data input. The first stage can be performed in various ways; in this paper, I employ full 3-D simulations for data generation. This stage can be very time consuming when considering all the complexity of the underlying resistivity structures, which requires fine-scale models and increases the computational cost. At the same time, data generation is perfectly parallelizable. Another thing to note is that the great success of DL and especially CNNs in computer vision has become possible due to the large public image repositories available. Creation of such repositories for geophysical data would lead to a substantial shift in the number and quality of DL applications in geosciences.

The approach to DL inversion presented here can be extended to other geophysical tasks, e.g., exploration. The main requirement to such inversion will be to perform well on new, previously unobserved inputs and not only on those which are similar to the examples used in training. The network does not need to be trained on all possible subsurface models (which will be impossible for any model having a practical amount of parameters). Instead, it should be provided with a sufficiently representative set of models and it will be able to learn the required mapping by itself. It also worth mentioning that a significant progress in NN accuracy can be achieved using transfer learning, i.e. copying the model and algorithm that is already known to perform best on another task that has been studied extensively. For example, a CNN trained on image classification can be used to solve other computer vision tasks. Several popular image segmentation networks \citep{long2015fully, badrinarayanan2017segnet} adapted the convolutional part of powerful classification networks and transferred their learned representations to the segmentation task. Modern deep networks may have many billions of parameters and data examples and, in some cases, even the entire trained network can be successfully applied to similar tasks.

\begin{figure}
\centering \includegraphics[width=0.63\linewidth]{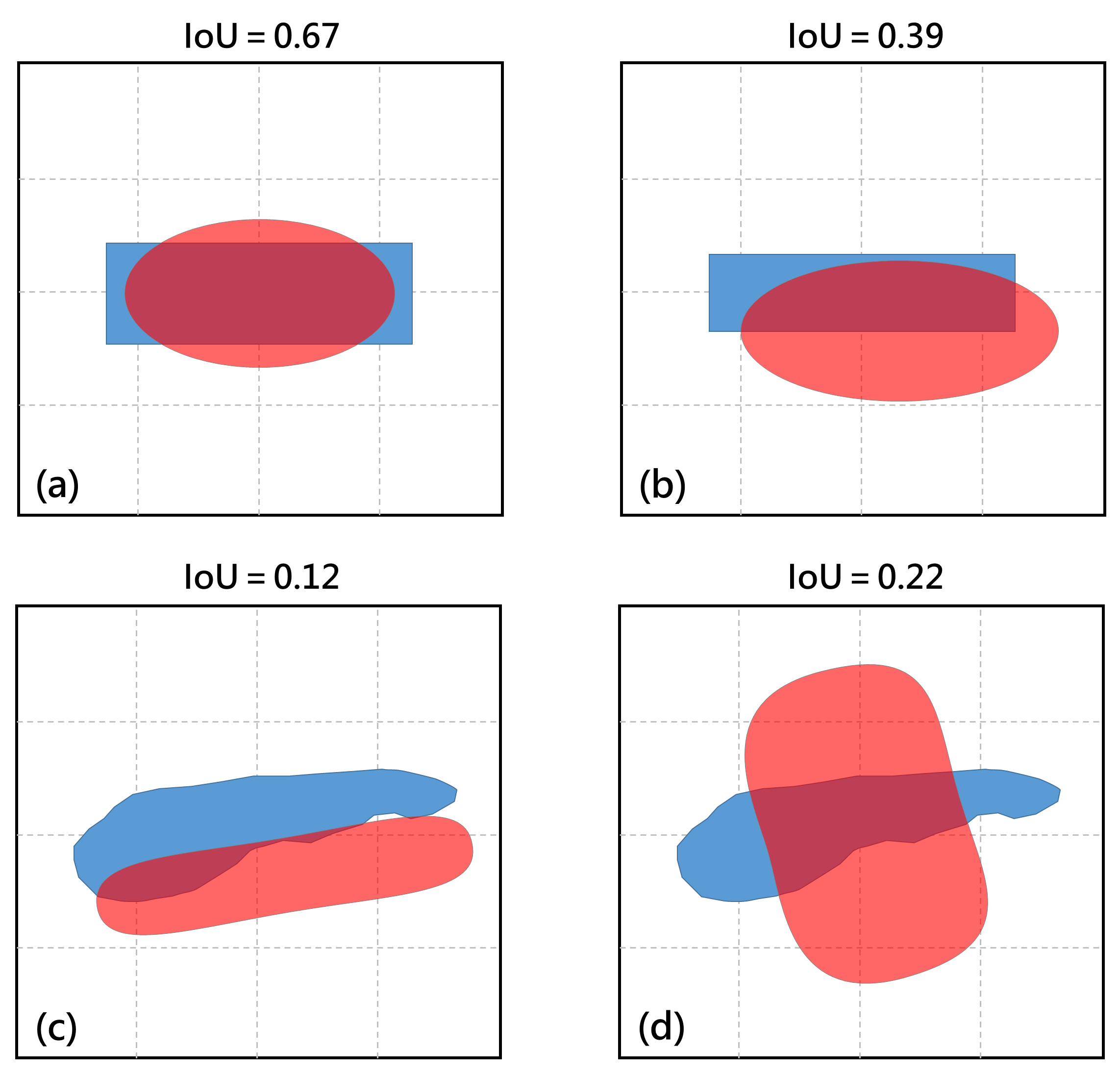}
\caption{Values of the IoU metric for different scenarios of anomaly detection. A good match between the true and predicted anomalies (a) has IoU of 0.67; a somewhat less precise match (b) has 0.39 of IoU score. Predicting shapes but slightly missing the locations (c) leads to a very low IoU of 0.12; at the same time, quite wrong prediction shown in (d) has a higher IoU of 0.22.}
\label{Fig:IoUExamples}
\end{figure}

The performance of the developed method can be improved by introducing new loss functions. IoU is a widely accepted evaluation metric in image segmentation but it might be not ideally suited to non-constrained EM inversion. Figure \ref{Fig:IoUExamples} illustrates some limitations of the IoU measure. The result shown in Figure \ref{Fig:IoUExamples}c would be considered good given the diffusive nature of the EM field in conductive media though its IoU is only 0.12, especially compared to the result shown in Figure \ref{Fig:IoUExamples}d (IoU $= 0.22$). New loss functions that take into account structural similarity of models have to be developed and tested. Sharpness of models, usually controlled by regularization in traditional inversion, is now determined by the training data. Networks trained on models with sharp geological features will learn to produce similar models and vice versa.

CNNs are not restricted to two-dimensional input topology and, as shown in Section \ref{sec:Ex2}, they can be applied to 1-D input data as well. Another type of NNs that is known to be highly efficient on sequential data is recurrent neural networks (RNNs). Output of the fully convolutional networks can also be extended to the third dimension. For example, a natural and rather straightforward extension of the proposed method is adding anisotropy or resistivity planes in the third dimension, thus allowing for 2.5-D or even 3-D inversion.

\section{Conclusions}

Significant recent developments in DL fueled by rapid advances in computing power have led to the widespread use of these methods in many domains of science and engineering. In this paper, I present a new method for EM inversion based on deep fully convolutional networks. The method requires a relatively small dataset for training and allows to estimate the resistivity distribution with high precision and orders of magnitude faster than conventional inversion methods. For generation of the training dataset, I perform a series of full 3-D simulations on synthetic models. Next, the electric and magnetic field components are shaped into 2-D or 3-D arrays and provided to the network to learn the mapping from the measured data to the model. The chosen architecture has no restrictions on the input dimensions and size. Hyperparameters for each particular network are selected guided by the validation error during the training stage. Both batch normalization and dropout are used to avoid overfitting on training examples and improve performance on new data.

Numerical examples for the borehole-to-surface CSEM configuration with a vertical electric dipole demonstrate the performance of deep NNs in 2-D inversion. By making use of sufficiently deep fully convolutional networks, the estimation of a subsurface resistivity model can be performed accurately and almost instantly. Training of the networks on modern GPUs can also be done in practical time using mini-batch adaptive learning rate algorithms such as Adam or Adadelta. Several examples ranging from $\text{CO}_\text{2}$ plume delineation to full resistivity inversion confirm the accuracy of the method. The quality of the DL inversion does not deteriorate significantly when random noise is added to the data. The full range of possibilities that DL potentially enables in EM inversion is significantly broader than described here. The future work will focus on applications of other DL methods such as recurrent and generative adversarial network architectures.

\appendix
\section{Data generation by fast forward modelling}
Both traditional deterministic inversion and data generation for NN training involve a high number of forward modelling simulations when Equation \eqref{eq:CurlCurl2} is solved for all values of the frequency parameter and source position. Thus, the total computational cost of these tasks is highly affected by the forward modelling efficiency. For a typical industrial-scale CSEM problem, one may have a number of transmitter and receiver positions in the range of several thousands. Computationally efficient simulations require a forward solver that can accurately and efficiently simulate the forward problem for multiple subsurface models.

The training data generation was performed using the parallel 3-D finite-difference code based on the curl-curl electric field formulation \citep{puzyrev2015review}. The code has hybrid MPI/OpenMP parallelization, which is well-suited for modern multicore processor architectures. Forward problems at different frequencies are processed independently on their computational nodes, so the parallel scalability is mainly limited by the size of the forward problems. For iterative solution of the resulting linear systems, I use PETSc parallel library \citep{balay2017petsc}.

ILU preconditioner is a rather common choice in EM modelling \citep{um2013efficient, ansari20143d}, which allows to reuse the precomputed factors. The models are sorted in a way that the difference between two consecutive models is minimal. The response of the previous model is used as an initial guess. Preconditioning reuse and use of initial guess resulted in $\sim$10x speed up compared to independent system solution. Parallelization was done over frequencies and different locations of the plumes since similarity of these models and hence the possible computational gain is small.

\begin{figure}
\centering \includegraphics[width=0.55\linewidth]{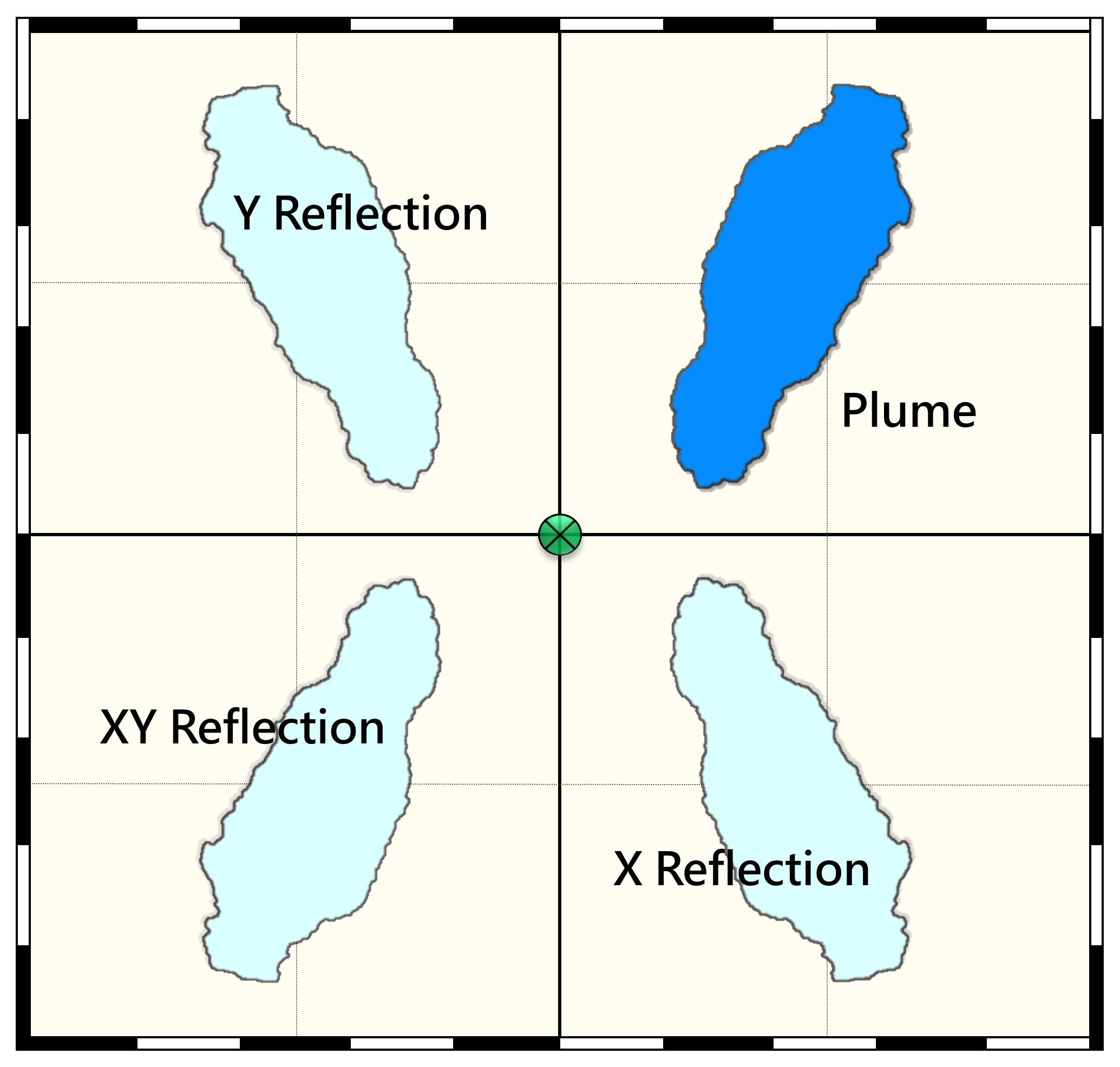}
\caption{Resistivity anomaly and its symmetric reflections. 2-D horizontal cross-sections through the middle of the storage aquifer where $\text{CO}_\text{2}$ is injected. The plume shape used in this example is taken from \cite{chadwick2010quantitative}.}
\label{Fig:SymmetricReflect}
\end{figure}

Figure \ref{Fig:SymmetricReflect} illustrates the data augmentation using the symmetry of the anomalous resistivity distribution in the 1-D background. Each anomaly has its X-, Y-, and XY-reflections, thus allowing for a fourfold increase of the initial dataset of 5,000 examples.

\section{Comparison of training algorithms}
Adaptive learning rate optimizers Adagrad, RMSprop, Adadelta, Adam, AdaMax, and Nadam have shown their efficiency for sparse datasets and does not require manual tuning of the step length. There is currently no consensus in DL community on which algorithm to choose, though Adam is often used as the default choice in many applications. Adadelta and RMSProp are essentially modifications of Adagrad that deal with its quickly diminishing step length. Adam adds bias-correction and momentum which makes this method more efficient at the end of optimization when the gradients become sparser. AdaMax is a variant of Adam based on the $L_{\infty}$ norm instead of $L_2$ norm, while Nadam is Adam with incorporated Nesterov momentum. For a more comprehensive overview of the theory, I refer the reader to \citet{ruder2016overview} and Chapter 8 of \citet{goodfellow2016deep}; the foundations of gradient-based optimization methods are given in \citet{nocedal2006nonlinear}. The field is rapidly changing and new methods and modifications such as AMSGrad and AdamNC are proposed every year.

\begin{figure}
\centering \includegraphics[width=1.0\linewidth]{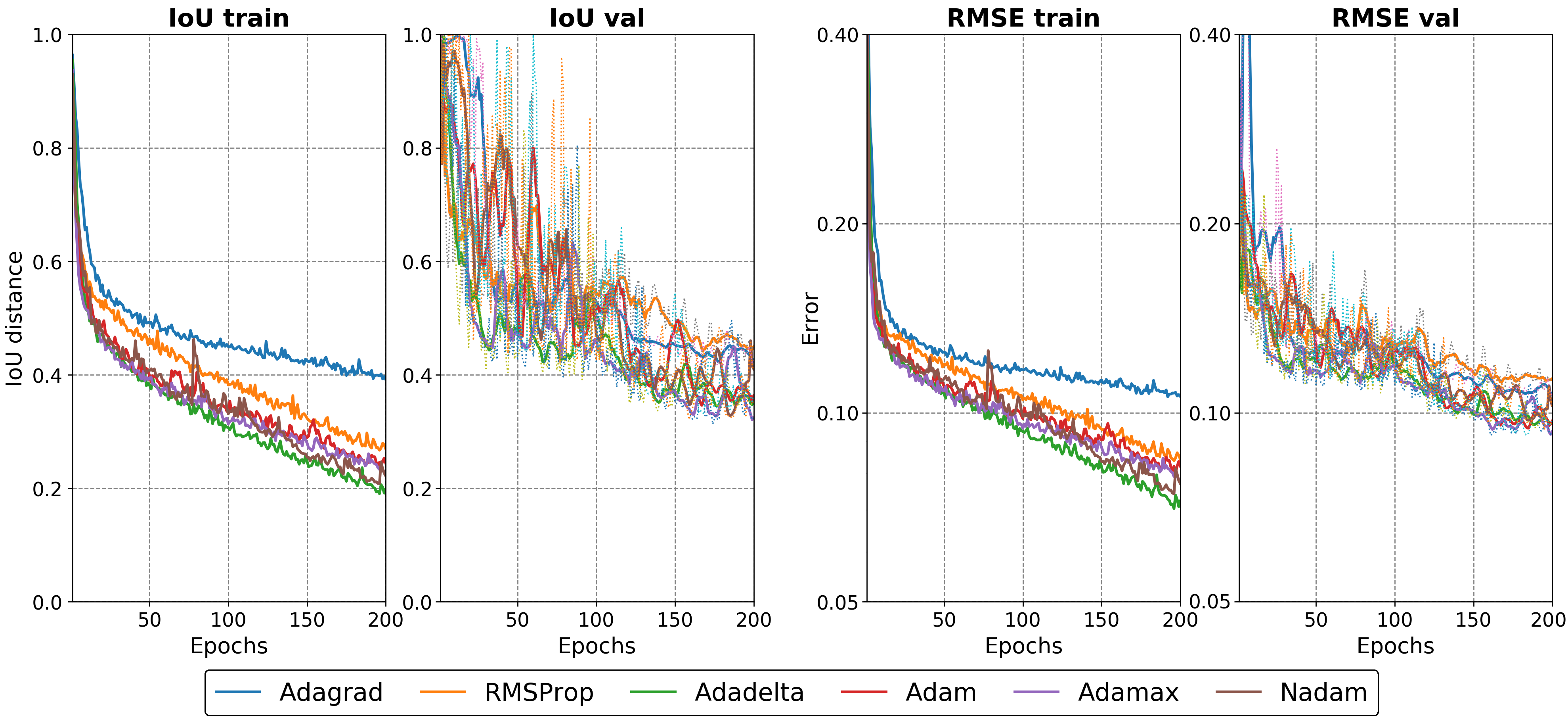}
\caption{IoU (left) and RMSE (right) training and validation errors of the optimization algorithms Adagrad, RMSprop, Adadelta, Adam, AdaMax, and Nadam using their default parameters from the Keras library \citep{chollet2015keras}. Due to their high oscillations, the validations errors are smoothed using a 7-point sliding window (thick solid lines); original validations errors are also shown in thin dotted lines.}
\label{Fig:OptimizerComparison}
\end{figure}

Figure \ref{Fig:OptimizerComparison} compares the performance of all standard adaptive learning rate algorithms for training a 2D CNN network with two convolutional blocks at each layer, 64-256 filters and 1.7 million parameters on the data of Example 1. Adagrad and RMSprop are the worst in minimizing the training and validation error, respectively. Adadelta has the lowest training error and performs well on the validation dataset. Adam and its variants also show high efficiency on both sets. Based on my personal experience with convolutional and recurrent network training, I recommend trying both Adadelta and Adam optimizers and picking the best one.

\end{document}